\documentclass[]{iopart}

\usepackage{textcomp}

\usepackage{graphicx}

\usepackage{epsfig}

\usepackage{fixme}

\renewcommand{\contentsname}{}

\begin{document}

\title{The Electronics and Data Acquisition System of the DarkSide
  Dark Matter Search}  

\author{ The DarkSide Collaboration; P. Agnes$^{1}$,
  T. Alexander$^{2}$, A. Alton$^{3}$,
K. Arisaka$^{4}$, H. O. Back$^{5}$, B. Baldin$^{6}$, K. Biery$^{6}$,
G. Bonfini$^{7}$, M. Bossa$^{8}$, A. Brigatti$^{9}$
% ...
}

\address{ (1) APC, Universit\'e ParisDiderot, Sorbonne Paris Cit\'e, 
75205 Paris, France;}
\address{(2) Physics Department, University of Massachusetts,
  Amherst, MA 01003, USA; } 
\address{(3) Physics and Astronomy Department, Augustana College,
  Sioux Falls, SD 57197,  USA;}
\address{(4) Physics and Astronomy Department, University of California, Los
Angeles, CA 90095, USA; }
\address{(5) Physics Department, Princeton University, Princeton, NJ
  08544, USA;} 
\address{(6) Fermi National Accelerator Laboratory, Batavia, IL 60510, USA;}
\address{(7) Laboratori Nazionali del Gran Sasso, 67010 Assergi, Italy;}
\address{(8) Gran Sasso Science Institute, 67100 L'Aquila, Italy;}
\address{(9) Physics Department, Universit`a degli Studi and INFN,
  20133 Milano, Italy;} 
\address{(10) Physics Department, Universit`a degli Studi Roma Tre
  and INFN, 00146 Roma, Italy;} 
\address{(11) Physics Department, Universit\'a degli Studi and INFN, Genova, 
16146, Italy; }
\address{(12) Kavli Institute, Enrico Fermi Institute, and
  Department of Physics, University of Chicago, Chicago, IL 60637, USA;} 
\address{(13) Skobeltsyn Institute of Nuclear Physics, Lomonosov Moscow State
University, Moscow 119991 ;}
\address{(14) Physics Department, Universit`a degli Studi Federico
  II and INFN, 80126, Napoli, Italy;} 
\address{(15) Department of Physics, University of Houston, Houston,
  TX 77204, USA} 
\address{(16) Petersburg Nuclear Physics Institute, Gatchina 188350, Russia;}
\address{(17) Physics Department, Universit\`a degli Studi and INFN, 
Cagliari 09042, Italy;}
\address{(18) Department of Physics and Astronomy, University of Hawai'i, 
Honolulu, HI 96822, HI;}
\address{(19) Joint Institute for Nuclear Research, Dubna 141980, Russia;}
\address{(20) Institute for High Energy Physics, Beijing 100049, China;}
\address{(21) IPHC,19)  Universit\'e de Strasbourg, CNRS/IN2P3,
  67037 Strasbourg, France; } 
\address{(22) School of Natural Sciences, Black Hills State
  University, Spearfish, SD 57799, USA;}
\address{(23) Physics Department, Virginia Tech, Blacksburg, VA 24061, USA;}
\address{(24) Institute for Nuclear Research, National Academy of Sciences of 
Ukraine, Kiev 03680, Ukraine;}
\address{(25) Physics Department, Temple University, Philadelphia,
  PA 19122, USA;} 
\address{(26) SLAC National Accelerator Laboratory, Menlo Park, CA
  94025, USA;} 
\address{(27) National Research Center Kurchatov Institute, Moscow
  123182, Russia;} 
\address{(28) National Research Nuclear University MEPhI (Moscow
  Engineering Physics Institute), 115409 Moscow, Russia;}  
\address{(29) Department of Chemistry, Biology and Biotechnology, Universit`a
degli Studi and INFN, 06123 Perugia, Italy;}    
\address{(30) Lawrence Livermore National Laboratory, 7000 East
  Avenue, Livermore,CA 94550, USA;} 
\address{(31) Smoluchowski Institute of Physics, Jagiellonian
  University, 30059 Krakow, Poland;} 
\address{(32) Physics Department, University of California, Davis,
  CA 95616, USA;}

\begin{abstract}
 This paper reports on the electronics and data acquisition 
system of a dark matter (DM) search using a 50 kg 
dual-phase, liquid argon time projection chamber (TPC) which was embedded 
in water and liquid scintillator veto detectors (DS-50). If DM 
is a subatomic particle, a possible candidate is a Weakly 
Interacting Massive Particle (WIMP), and the DS-50 experiment is 
a direct search for evidence of WIMP-nuclear collisions at the 
Laboratori Nazionali del Gran Sasso (LNGS). The light 
from nuclear excitation in the TPC as a result of 
a WIMP-Nuclear collision, is collected by 38 
photomultiplier tubes (PMT) positioned on the top and bottom of the 
cylindrical TPC cryostat. The two veto detectors, instrumented with PMT, 
shield the TPC and tag background events due to 
cosmogneic and local radioactivity. All PMT signals from both the TPC and 
Veto systems are 
digitized and their waveforms transferred in parallel 
to event building computers running the ARTDAQ software developed at 
the Fermi National Laboratory (FNAL). This paper describes the 
system triggers,  the synchronization of the data flows, and the event
building for both the TPC and Veto.
\end{abstract}


\maketitle
 \newpage
    
\maketitle

\section{The DarkSide Experiment}

A wide range of astronomical evidence suggests the existence of a 
gravitationally-interacting, non-luminous component of the universe which 
has yet to be identified. This Dark Matter (DM) component  comprises 
approximately 27\% of the energy density of the universe, and is 
responsible for the galactic structures visible today ~\cite{astrophy}. 
There are viable arguments that 
DM could be composed of Weakly Interacting Massive Particles (WIMP), and the 
DarkSide-50 (DS) experiment is a search for  
WIMP-nuclear collisions as a fixed detector on Earth moves 
through the WIMP field of the Milky Way. 
Such interactions are rare, and impart usable recoil energies of
$<100$keV to a nucleus \cite{goodman}. 
Thus to reach the required sensitivities, detectors 
having large target masses and background suppression are needed. A number 
of technologies 
are now used for direct detection of dark matter WIMPs. These include;
cryogenic bolometers with ionization or scintillation detection,
 sodium/cesium iodide scintillation detectors, bubble chambers,  
a point contact germanium detector, and liquid noble gas detectors
filled with Xenon or Argon.  \\

\begin{figure}[h]
  \hfill
  \begin{minipage}[b]{4.5 cm}
    \begin{center}  
      \epsfig{file = 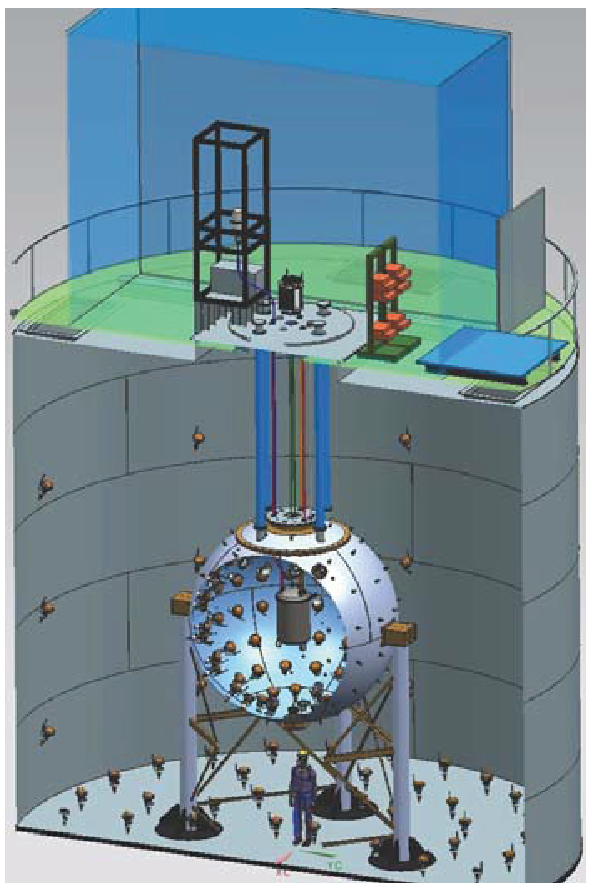, width= 5.0cm, height=8.0cm}
    \end{center}
  \end{minipage}
    \begin{minipage}[b]{6.5 cm}
    \begin{center}  
      \epsfig{file=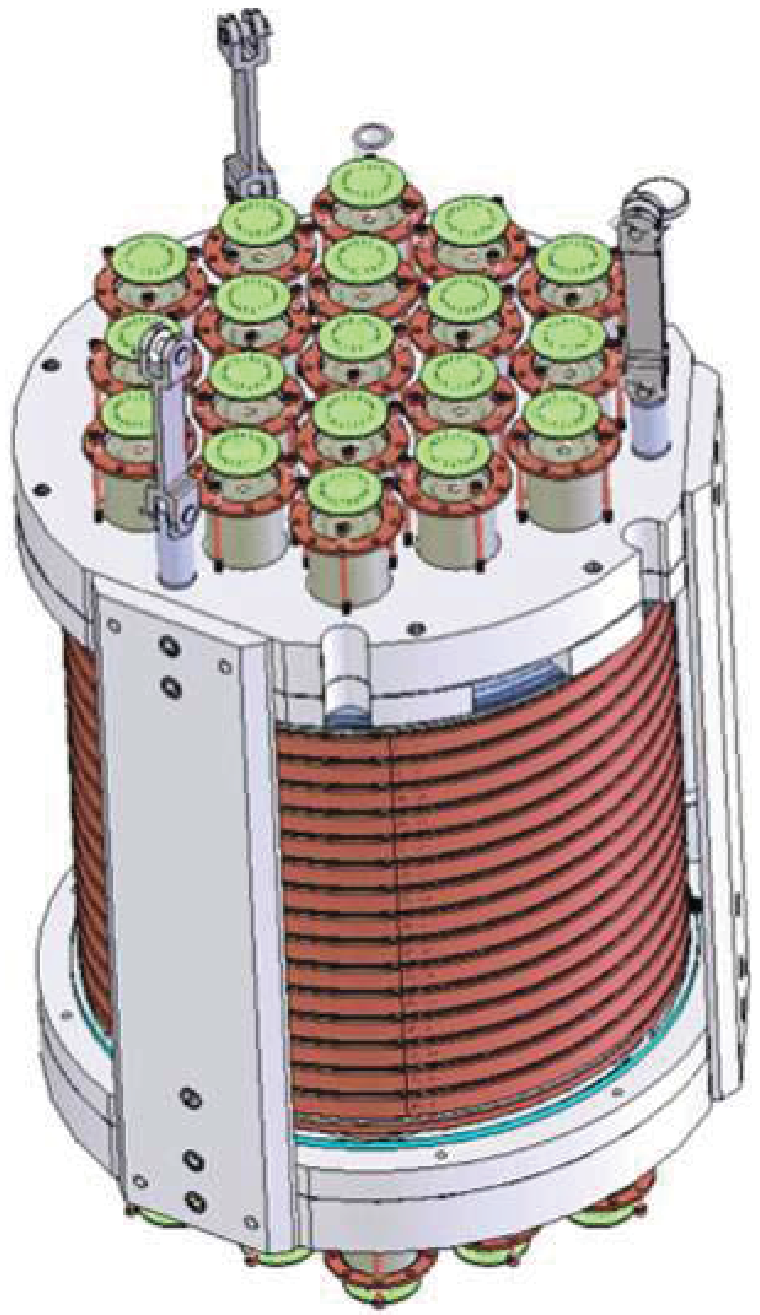, width=5.0cm, height=8.0cm}
    \end{center}
    \end{minipage}
  \begin{minipage}[t]{5.5 cm}
  \caption{\sloppy The DarkSide experimental layout showing the water and
scintillation vetoes, the TPC cryostat, and the clean room (CLR) atop the CTF}
  \label{dsexp}
  \end{minipage}
    \begin{minipage}[t]{5.5 cm}
  \caption{The DarkSide TPC showing the PMT and field shaping rings. The TPC is immersed in Lar but has a gas pocket just below the upper PMT window.}
  \label{dstpc}
  \end{minipage}
  \end{figure}

The heart of the DarkSide-50 (DS) experiment 
is a 50 kg Time Projection Chamber (TPC) which is a two-phase liquid
argon (Lar)  
detector. The DS system is designed to achieve background-free 
operation for at least 0.1 
tonne-year of exposure. The TPC exploits
the extraordinary pulse shape discrimination (PSD) of LAr ~\cite{pulse_shape}
providing powerful rejection of electromagnetic (electron or 
gamma induced) signals. Cosmogenic 
background is suppressed to very low levels by mounting DS at a 
 3800 mwe (water equivalent) depth at the  Laboratori Nazionali del 
Gran Sasso (LNGS) laboratory ~\cite{bxmuonflux} ~\cite{bxcosmogenic}, and 
surrounding the TPC with
two nested veto shields of high purity water (CTF) and liquid scintillator 
(LSV), 
Figure ~\ref{dsexp}. In addition, local radioactive background 
is significantly reduced 
by efficient chemical and cryogenic purification methods
\cite{purification}.  The experiment uses  
argon obtained from gas wells (Uar) which is depleted in the 
radioactive $^{39}$Ar isotope ~\cite{Uar}.  
Despite the small detector
size, DS has a sensitivity to WIMP-nucleon cross sections of approximately 
$10^{-45}$ cm$^{2}$ for a WIMP  mass of $100$ GeV/c$^2$. \\

This paper describes the DS electronics, 
the data collection and event-building systems (DAQ), and the slow control and 
data monitoring systems of the experiment. \\

\section{Detector Systems}

DarkSide-50 consists of three embedded detector systems.  Viewed from the inner 
to the outer these systems are: a Lar TPC, a liquid scintillator system (LSV) 
composed of 50\% PPO (2,5-Diphenyloxazole) + 50\% TMB
(Trimethylbenzene) ~\cite{lsv_scint}, and a water 
Cherenkov Veto system (CTF) which also acts to shield the LSV and
TPC from local radioactive background. An enclosed clean 
room (CLR) sits atop the CTF, and is used to insert the TPC, and to
provide internal access to the vetoes  when the water and
scintillator are removed. The CLR isolates the TPC 
and its components from surface radioactivity, 
particular radon, during assembly.\\ 

Electrical connections to the TPC, approximately 5-10~m in length,
are made through vacuum-tight pipes from the TPC to a feed-through in a 
flange on the 
floor of the CLR.  To minimize noise from the PMT 
signals, the TPC front end electronics are located in the CLR racks 
just above the flange, which is as close as possible to the
TPC. However, readout 
of the veto signals from the LSV and CTF are patched through the 
side of the CTF to a control room (CoR) by cables some 40~m in length. Thus 
synchronization of these systems is one of the electronic challenges 
and is addressed below. \\     

\section{The TPC}

The TPC, shown in Figure ~\ref{dstpc}, is a cylindrical detector 
containing 50~kg of 
Lar. The inner cylindrical surface of the detector is a 
Teflon reflector coated with wavelength shifter  (tetraphenyl 
butadiene) ~\cite{tpb}. The coating 
absorbs the
128~nm scintillation photons emitted by excitation in the Lar, 
and re-emits visible 
photons which are viewed by thirty-eight Hamamatsu low-background
R11065 PMT ~\cite{hamm11065}.  The PMT are 
divided equally between the top and bottom 
surfaces of the cylindrical TPC, and receive the
 wavelength-shifted light through
fused silica windows. These windows are also coated with the wavelength
shifter on the inner surfaces and have transparent conductive layers of 
indium tin oxide (ITO) evaporated on both the inner and outer surfaces. 
Coating both surfaces with ITO allows
the inner window surface to serve as a grounded anode (top) and a high voltage 
cathode (bottom) of the TPC, while maintaining the outer surfaces
at the average PMT photocathode potential. The fused silica anode 
(top) window has 
a cylindrical rim extending downward into the TPC to
form a 1 cm-thick gas pocket above the Lar. High voltage is applied 
between the TPC cathode and anode to produce, by field shaping rings, 
a uniform electric field throughout the active volume. 
Also, light from Lar excitations (S1) 
 is collected by the PMT.  Electrons from ionization drift
upwards in the electric field, eventually reaching the 
liquid surface where they are extracted, and 
accelerated in the gas pocket producing secondary ionization light (S2). 
The drift time of the ionization charge in the Lar, and the amplitude 
of the light signal in the gas are adjustable by 
varying the potentials on the cathode and an extraction 
grid placed just below the Lar surface. This design not only provides
three-dimensional information of the 
position of the event in the Lar, but also
allows additional discrimination between the ionization due to Ar recoils and
that of electromagnetic interactions, ({\it ie} $\beta$ or $\gamma$
events).  However, the experimentally applied discrimination mainly 
uses the ratio of the 
fast component of the (S1) ionization signal (the first 90 ns) to its total 
integral ~\cite{pulse_shape}. Thus, it is important to minimize   
electronic noise and use a low frequency cutoff to accurately
determine the timing of the detected photo electrons. \\

\subsection{TPC Front-End Electronics}

As explained above, wavelength-shifted light from an event in the Lar was 
viewed through fused silica windows by 38 low-background PMT. 
Although the PMT were not placed in the sensitive volume of the TPC, 
 they were immersed in LAr and operated at a temperature of 
approximately 
87~K. After extensive testing, it was found that charge accumulation on
internal components in the PMT induced erratic behavior.  This was 
mitigated by decreasing the tube voltage, reducing the PMT gain to a 
few $\times \, 10^{5}$. However, the reduced gain required a local 
pre-amplifier operating at Lar temperatures to drive the 
signals through the approximate 5-10~m of cable to the amplifiers in the CLR 
without 
significant contribution from noise and dispersion. Both the TPC 
amplifiers and digitizers were located in racks in the CLR, just 
above the CTF. \\

\subsubsection{Pre-amplifier}

\begin{figure}[t!]
  \hfill
  \begin{minipage}[b]{4.5 cm}
       \epsfig{file =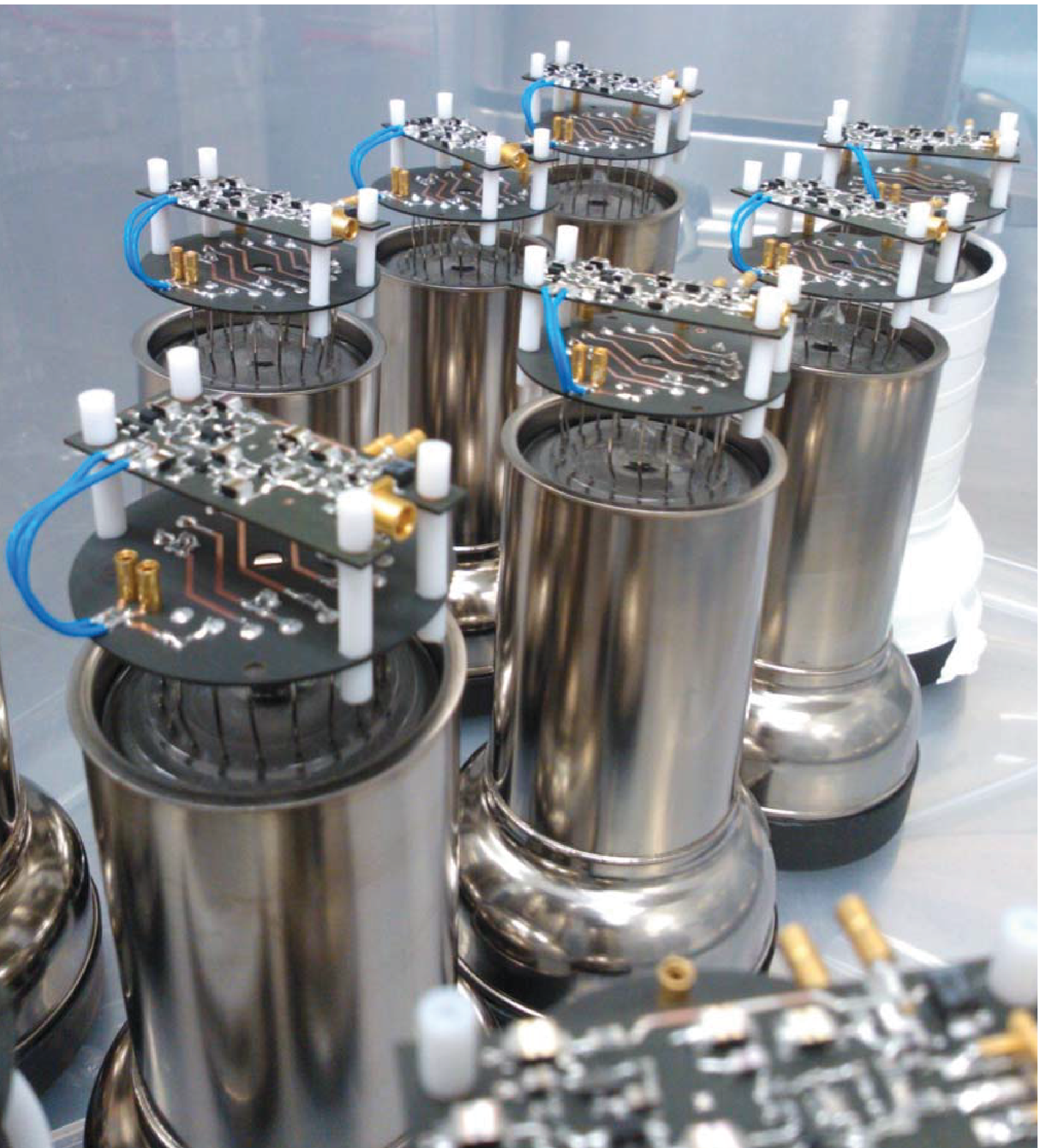, width = 4.5 cm, height = 4.5 cm }
\begin{center}
       \caption{\label{preamp_fig}}
\end{center}  
A picture of the cryogenic pre-amplifier 
   mounted of the PMT base
 \end{minipage}
  {\hspace{1. cm}
  \begin{minipage}[b]{5cm}
   \epsfig{file =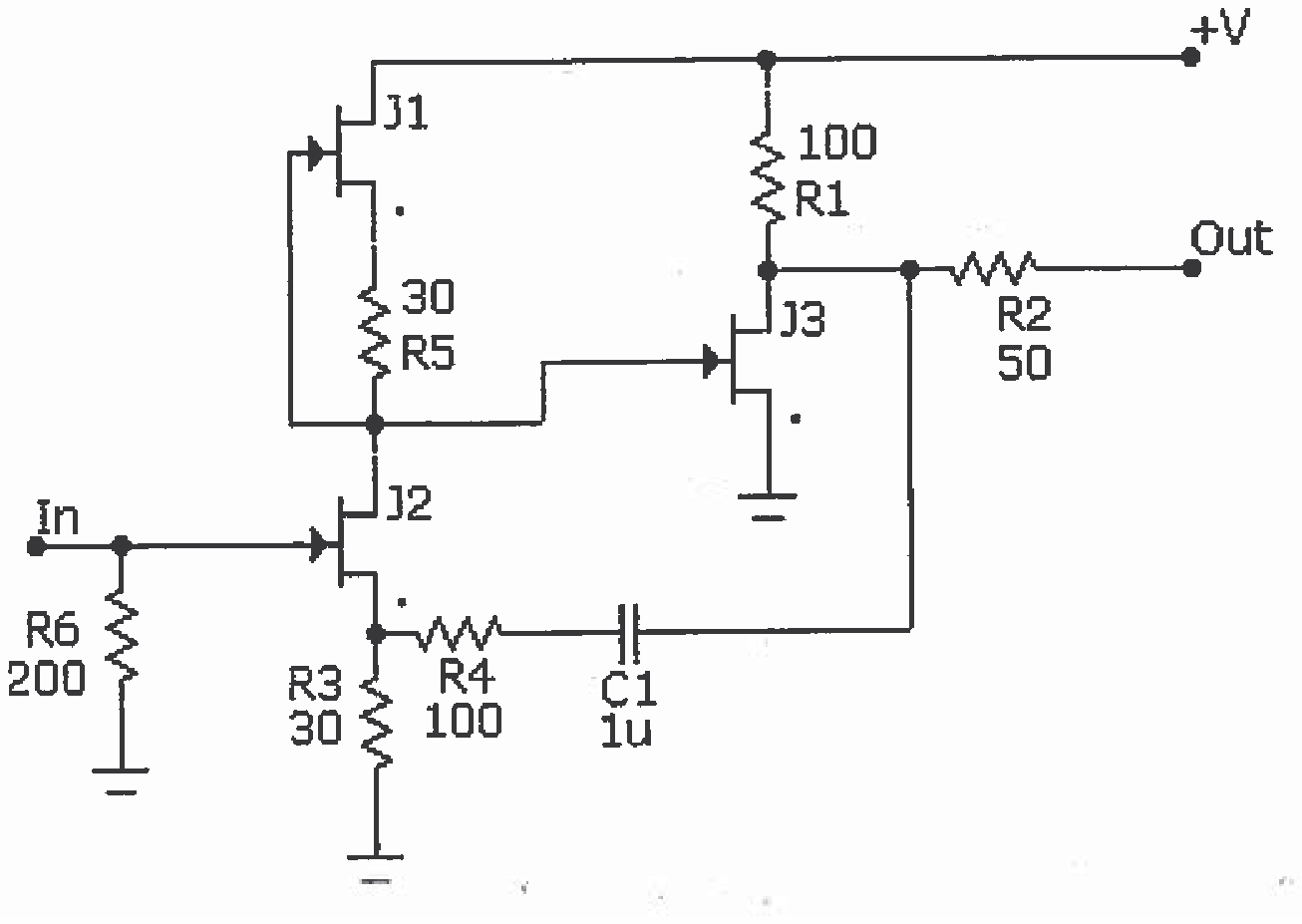, width = 6. cm, height = 6. cm }
\begin{center} 
  \caption{\label{preamp_circuit}}
\end{center} 
 The circuit diagram of the 
cryogenic pre-amplifier 
\end{minipage}}
\end{figure}

The cryogenic pre-amplifier is mounted directly on the PMT base as
shown Figure ~\ref{preamp_fig}. It was constructed using discrete components 
as illustrated in  Figure ~\ref{preamp_circuit}. 
Power-loss density is important as the pre-amplifier is mounted on the PMT 
base which is immersed in Lar. Local boiling can cause a multitude 
of problems in high voltages, electronics, and TPC operations. 
Consequently the pre-amplifier is designed with a total power
consumption of less than 90 mW/channel. It is also 
constructed to be as radio-pure as possible, which required the use of a Cirlex 
~\cite{cirlex} circuit board, low radioactive components, and the avoidance of 
copper-beryllium (CuBe) connectors.
Its input impedance is determined by the internal stray-capacitance
of the PMT and the optimization of signal-to-noise by waveform
shaping.   The relatively 
high-load resistance produces a passive gain with respect to the PMT 
coupling and is back terminated by 50 $\Omega$.  The active gain of 
the pre-amplifier is 3 V/V, for a total 24V/V compared to the output 
without a pre-amplifier.  The maximum output 
is 3V providing a dynamic range in excess of 1500 photoelectrons (pe). 
Pole-zero cancellation is applied to the output signals sent  
to the amplifier. High quality, double shielded cables 
~\cite{multiflex} are used to transmit these signals from the TPC to the 
amplifiers in the CLR.  \\

\subsubsection{Noise Abatement}

Special care was taken to maintain high signal integrity during the 
implementation.  Thus, SMA 
RF connectors and double shielded
signal cables are used where ever possible. The cryogenic
pre-amplifiers used surface mounted components with a grounded, full-copper
backplane. The power supply was carefully filtered to
reduce noise components. Impedance matching and signal reflection of the 
output signals were also adjusted to minimize cross-talk and
electromagnetic interference (EMI).\\

While the hermetically sealed cryostat containing the
pre-amplifiers should provide shielding from external noise sources, 
penetrations by signal, test, and power cables conveyed external noise to
the pre-amplifiers. This is exacerbated by the requirement that the PMT
anodes were maintained at ground potential in order to use DC signal
coupling, so that the metal envelope of the PMT remained floating
at the cathode potential and could not be easily shielded. 
Consequently interference pick-up by the PMT dynodes
becomes impressed on the photocurrents prior to amplification. 
Given the above issues, noise abatement required simultaneously 
considering the full detector system, including the cryogenetic and the 
slow control components, as well as the TPC electronics. It was 
found that pump-drivers, switching power supplies,
telecommunication transceivers, and high voltage supplies 
all provided contribution to the signal noise.  These were addressed
in situ after installation. \\

Pump motor controllers were especially noisy sources because most sent
unfiltered and un-shielded signals to the appliance they controlled.
These sources of noise were mitigated by variac-based sinusoidal motor
controls.  Noisy switching power supplies were identified and
individually treated by filtering or replacing the supplies with linear 
components. In particular, noise from the PMT high voltage (HV) 
supplies was reduced by adding filters to the HV output. \\

Ground loops were also mitigated by floating all
connections through the cryostat wall. At the
same time, all cable shields were connected via low-inductance AC coupling
to the cryostat surface, to reduce the RF energy which would otherwise
enter the cryostat. As a result, the noise at the output of the
amplifier chain is dominated by the pre-amplifier electronic noise 
floor.  Thus there are no
noise-dominant spectral components within the frequency range of interest.  The
output noise is  70 $\mu$V for an active bandwidth of 150 MHz. \\

\subsubsection{Amplifier}

\begin{figure}[h]
\begin{center}
\includegraphics[width = 11cm ]{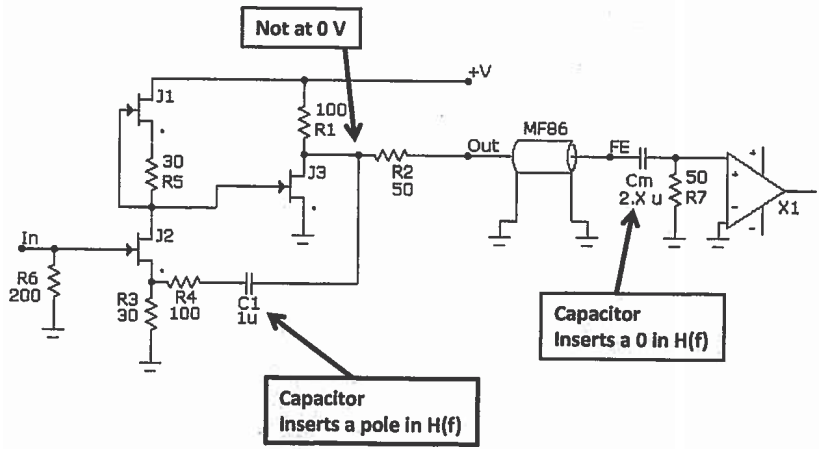}
\caption{The coupling circuit between the pre-amplifier and the amplifier}
\label{fe_coupling}
\end{center}
\end{figure}

The amplifier in the CLR provides further amplification and signal 
conditioning. The coupling between the 
pre-amplifier and the 
amplifier boards is illustrated in Figure ~\ref{fe_coupling}.
One amplifier module handles five channels and provides the 
following outputs for each channel. \\

\begin{itemize}
\item an amplified output with gain of 10 V/V and an input noise 
equivalent of 20 $\mu$V over a bandwidth of 200 MHz. This output is typically 
used for monitoring the performance of specific channels
\item a shaped output with a gain of 10 V/V with a bandwidth of 90 MHz and 
a noise equivalent of 15 $\mu$V. The bandwidth is matched to the 
sampling frequency of the CAEN V1720 digitizer which samples at 250 MS/s
\item an attenuated output with gain of 0.5 V/V with bandwidth of 
40 MHz. The bandwidth of this output 
is matched to the sampling frequency of the CAEN 1724 digitizer
sampling at 100 MS/s
\item discriminated low-voltage-differential-signals (LVDS), with a 
minimum pulse duration of 4 ns. 
The discrimination threshold is remotely programmable.  These signals 
are sent to the trigger unit, and to scalars for monitoring
\end{itemize}
 
All outputs are tuned to match the  -2 to 1 V dynamic range of the 
digitizers. The typical output noise 
is approximately 1~mV, 50\% of which is due to the digitizer.      

\subsubsection{Digitization}

The $\times$10 gain signals are sent to CAEN V1720 250 MHz 12 bit 
waveform digitizers \cite{caen172x} and the low gain signals sent to 
CAEN V1724 100 MHz 14 bit 
digitizers. Both digitizer types are one unit 6U VME64X VME modules, servicing
eight channels. They have circular memory buffers of 10 Ms/ch, and are locally 
FPGA compliant.  The
use of two digitizer types extends the dynamic range, providing a linear 
response between 1 and 10,000 pe, capturing signals without 
saturation from both Lar and Ar gas. The large 
dynamic range
also improves calibration options and provides a more detailed 
study of backgrounds.\\

The multiple ADCs are synchronized with a common 50 MHz clock and external 
trigger.  They run in parallel and 
require three common control signals: a ``Trigger'', a ``Run-enable'',
and a ``Clock''.
 The Trigger and Run-enable can be either TTL or 
NIM level signals, however the clock signals are differential LVDS. 
The distribution of all signals requires equal delays in a 
star-like topology in order to keep timing synchronization. \\

\section{Veto Systems}

As described above, there are two embedded veto systems, 
a water Cherenkov detector (CTF), and a 
liquid scintillator detector (LSV). The effectiveness of these veto systems 
is described in ~\cite{ae_evh} and included references. The CTF is an 
active veto, tagging penetrating cosmic muons and  
cosmogenically produced charged showers in materials surrounding 
DS and shielding 
the TPC and LSV from external radioactivity. The CTF is a cylindrical, 
steel tank, 11~m in diameter and 10 ~m high, filled with 1000 tons of 
high-purity water. 
It is instrumented  along the floor and bottom half of the cylindrical wall 
by 80 ETL-9351 8 inch PMT \cite{etl9351}. 
Studies revealed little spatial dependence of the PMT 
placement, so they were more-or-less equally spaced. The  
walls are covered 
with reflecting ~\cite{tyvek} Tyvek  sheets. The veto electronics for 
both the LSV and CTF are located in the CoR.\\

The LSV is a 4~m diameter, stainless steel sphere which lies inside 
the CTF and
encloses the TPC.  It is filled with 
liquid scintillator and instrumented with 110 Hammatsu 
R5912-HQE-LRI 8 inch PMT ~\cite{ham5912}. The interior walls of the 
sphere are covered 
with reflecting Lumirror  sheets \cite{lumirror}. 
The LSV is designed to moderate, capture, and thus provide a veto
signal for neutrons  
which might enter, or exit, the TPC.
Neutrons can be thermalized by scattering from protons in the
scintillator liquid,  
and are efficiently captured by $^{10}$B nuclei. 
Capture on $^{10}$B proceeds to the $^{7}$Li ground state (gs) and a $^{7}$Li
excited state with branching ratios of 6.4 \% and 93.6 \%,
respectively. The ground state decays by emitting a  $1775$\,keV
$\alpha$ particle, and the excited state decays with emission of 
a $1471$\,keV $\alpha$ particle and a 478 keV $\gamma$.\\

The measured LSV light yield is $(540 \pm 20)$\,pe/MeV.
Scintillation light produced by the nuclear products of the 
$^{7}$Li(gs) decay is quenched and is expected to have a
beta-equivalent energy between 60 to 60 keV corresponding to 25-30 pe. 
Neutrons can also capture on hydrogen with the emission of a $2.2$ MeV
$\gamma$-ray yielding approximately $1100$~pe.\\

Most photons are collected in about 100~ns, but the tail of the distribution 
extends  some 300~ns, and must be considered  
when designing the appropriate
dynamic range of the acquisition electronics. 
For example, the average number of photons per channel in an event 
with energy $< 400$\,keV is typically 
less than one, 
but can increase by more than an order of magnitude when a 
neutron is captured near a PMT. In addition, a muon crossing the LSV 
produces a large signal  
corresponding to approximately 2~MeV per transit-length in the scintillator.
Thus to correctly determine the path of a muon, the veto electronics
must be able to efficiently detect a signal 
as small as 0.3~pe, but should
gracefully handle signals which saturate the front end electronics. 
Although the most
 important events occur at low energy, the system should recover 
from saturation in less than 2~$\mu$s. \\

Good timing is also important in order to obtain track reconstruction.  
This requires knowledge and stability of
 the relative time differences between PMTs for both vetoes
~\cite{borexino_1}. Therefore 
time alignment 
and synchronization among channels must have a precision of $\le$ 0.3~ns, 
and remain stable for at least a 24 hour period.\\

\subsection{Veto Front-end Electronics}

\begin{figure}[h]
\begin{center}
\includegraphics[width = 9cm ]{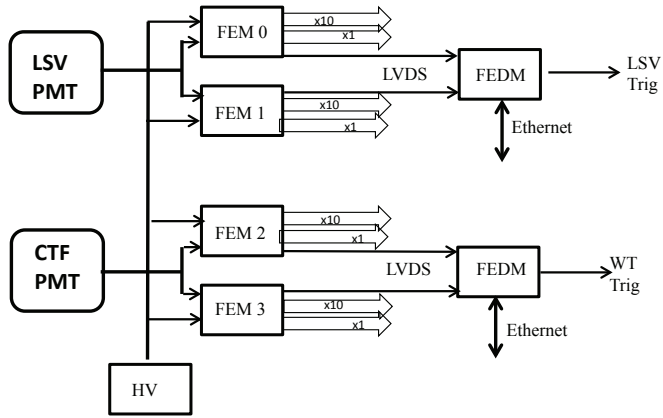}
\caption{A block diagram of the Veto electronics system}
\label{veto_circuit}
\end{center}
\end{figure}

Output signals from the veto PMTs are AC coupled directly to 
amplifiers through approximately 40~m of low dispersion, 50~$\Omega$, coaxial 
cable. The front end analog 
boards (FEAB) and 
front end digital boards (FEDB)
are designed to match the input timing, noise, and impedance of 
the PMT signals, but more importantly they match the bandwidth of the waveform 
digitizers. One PCB (board) contains 16 channels arranged in 
2 rows of 8 channels, and has 8 layers  with dimensions 274$\times$274~mm. 
Figure ~\ref{veto_circuit} is a schematic diagram of signal flow from the PMT 
through the FEAB to the FEDB.\\

The FEAB configuration uses 3 Texas Instruments operational
amplifiers per channel ~\cite{TI3201, LMH6559}, one with a gain of 1 
and two with a gain of 10. The x10 signals are
split providing digitizer inputs, monitors,
discriminated outputs,
and a summed signal for 16-channels. The circuit has 230~MHz bandwidth 
and a total noise of 200~$\mu$V$_{rms}$.  A linear power 
supply, able to deliver 15~A with a dual voltage $+/-$7.5~V, was
constructed to mitigate signal noise induced by the power source. 
Discrimination uses a dual, voltage comparator ~\cite{compar} with 
differential output.  The signals are sent by LVDS to the digital
board. The discrimination threshold is set using a standard by I$^2$C
bus with an 8 bit DAC.
 The rear panel of the FEAB 
has 16 HV connectors which connect the PMT cables directly 
to the CAEN A1536 High 
Voltage Boards ~\cite{caenhv}, and these are housed in a CAEN 
mainframe ~\cite{caenmf}.\\

\subsubsection{Veto Digital Boards}

The FEABs are controlled by a  FEDB.
These boards contain a Xilinx Spartan 6 FPGA ~\cite{spar6} which is used 
to generate a veto trigger.  The trigger firmware is constructed from  
logical patterns in the discriminator inputs from the FEAB. 
These input signals arrive at a mezzanine board on the FEDB where 
they are duplicated, one sent to a time to digital converter (TDC) and
the other converted to the single-ended TTL 
standard of the FPGA. A master clock signal is used for global
timing. The FPGA is interfaced to a PIC-32 micro-controller ($\mu$C) 
~\cite{micro_controller} via an I/O bus  
controlled by the $\mu$C. Both the FPGA and $\mu$C firmware
were developed for the project.\\

Some slow control operations are also handled by the FEDB, and
these are managed by the $\mu$C using Ethernet 
and FPGA connections.
The $\mu$C runs a single process which obtains an IP address from the 
LAN DCHP and acts as a command interpreter, waiting for incoming 
socket connections on a specific port.
In addition to resolving commands, the $\mu$C 
 dialogues via I$^2$C to set and read the 12 bit channel input 
offset-compensations and the 8 bit discriminator thresholds. 
It also controls an alphanumeric display via an I$^2$C bus 
to display data and status of a specific channel.\\

\subsubsection{Digitization}

The PMT signals from both the CTF and LSV
are amplified by the FEAB boards and their waveforms are digitized by 
commercial National Instruments PXIe-5162 modules ~\cite{NI5162}.
The  x10 channels are placed in 4 PXIe chassis, with two
chassis used for the CTF and two for the LSV. 
Each chassis houses analog inputs for the waveform digitizers (56 for
 the LSV, 40 for the CTF) and connects to digital TTL lines for the trigger and
 synchronization interface. The waveform is sampled at $1.25$~GS/s
(800~ps/sample) with 10 bit 
resolution (2 byte). Each waveform digitizer has 4 BNC input channels and 
is sampled asynchronously and continuously. When a trigger is received,
 a block of samples within an acquisition window is stored in the 
digitizer memory buffer. The timing of the trigger with respect to this 
window is configurable, so that samples 
prior to a trigger can be stored. The input voltage range for each
channel can be selected as either $0.1$, $0.2$, $0.5$, 
$1$, $2$ or $5$ V peak to peak, with the usual amplitude range  
set between $0.1$ and $0.9$ Volt. ~\cite{micro_controller}\\

\subsubsection{Zero Suppression}

A zero suppression algorithm is applied  to
 reduce the quantity of stored data.  
This operation has a user specified threshold, 
 minimum  width, and number of collected pre and post-samples.
When the waveform crosses, and stays above a threshold for a 
specified number of samples (minimum width), the algorithm returns the 
entire waveform between the upward and downward threshold crossings 
including a specified number of pre and post-samples. If the 
waveform goes below and then 
back above threshold before the specified number of post-samples are collected,
the algorithm waits for the waveform to drop back below threshold again before
counting post-samples.
Therefore, if two pulses appear on the same channel with overlapping zero 
suppression windows, they are combined into one larger pulse.
All the samples outside the accepted sample pulse widths are discarded.
Data are usually taken with a zero suppression threshold of $-30$ or 
$-50$~mV, a minimum width of 4 samples ($3.2$~ns), and with pre
and post-sample widths set to 25~ns.\\

\subsubsection{Veto Controller}

The NI PXIe-8133 controller \cite{NI8133} is an Intel Core i7-820QM
quad-core processor ($1.73$ GHz frequency), with $3.06$\,GHz (single-core
turbo-boost) running  the Windows7 OS.  
It is housed in a NI PXIe-1075 chassis and connects to the PXIe
modules using the PXI Express system. 
Two gigabit Ethernet ports are used to connect the controller to the
local area network. 
The data acquisition code in the controller, asynchronously
collects data from the digitizers and performs zero suppression. The
zero suppressed waveforms are then transmitted in a packet to the veto
builder via TCP/IP.  
The controller also receives commands from the master run controller.\\

Each processor controls up to 14 digitizer boards, one NI PXIe-6674T
timing and synchronization module \cite{NI6674}, and one NI PXIe-7961R
FlexRIO FPGA module \cite{NI7961}. 
The veto system requires four PXIe
chassis for the 190 channels, with each chassis housing one PXIe controller. \\

\section{Trigger}

The TPC and veto systems can operate independently, but must be 
synchronized in order 
to correlate TPC events with veto signals. Although the
the  two systems are physically displaced, the main difference between
them is the width 
of the data acquisition window, which  
is $\ge$ 350~$\mu$s for a TPC event, while the veto has either 
a ``short window'' of
$6.5$\,$\mu$s or a ``long window'' of $70$\,$\mu$s. 
  TPC triggering is obtained by logically processing signals received 
from the front-end, to determine the number of pe within
a set timing window, and the veto trigger is generated in a
similar way.
Synchronization between the systems is obtained through a high precision
time-stamp obtained from  a common 50 MHz clock slaved to a 1 
pulse per second (1PPS) signal 
received from the LNGS GPS. The TPC trigger is always the
trigger-master when collecting synchronized data.
However, each system can be locally triggered for calibration 
and testing. \\

\subsection{TPC Trigger}

The TPC data acquisition system (DAQ) includes one VME crate containing
5 CAEN V1720, 5 CAEN V1724 modules, a CAEN V1495 
\cite{caenV1495} logic module, and a V976 NIM-to-TTL fanout.
When an analog signal in any of the 38 PMT 
channels of the TPC
crosses a threshold, a logic pulse of fixed width representing a pe(s)
is generated 
and sent to the CAEN V1495 logic module, which is the
trigger master. 
The TPC trigger in the V1495 is 
obtained by firmware processing any selected set of these 
digitized PMT ``hits'' by compiling a running 
count within a sliding time window. When the running count 
crosses a defined threshold, a trigger pulse is initiated and a
latch generated to inhibit further triggers until the active trigger 
is processed. The veto trigger is generated in the same way using the FPGA 
in PXIe-7961R modules.
The parameters used in the
 FPGA trigger firmware are listed in Table ~\ref{tpcfirmware}. 
Parameters marked with [+] can be modified by inputs through the VME 
interface to the V1495. The TPC V1495 trigger module is the trigger 
master for both TPC and Veto 
systems. It includes appropriate dead time in the data acquisition window for 
either LSV or CTF Veto trigger types in order to prevent trigger overlap.  \\

\begin{table}[h]
\begin{center} 
\caption{ Parameters used in the TPC Logic unit to determine a trigger. 
Items market with a [+] can be modified by the VME interface to the logic unit}
\label{tpcfirmware}
\begin{tabular}{ll}
\hline
\multicolumn{1}{c}{Parameter} & \multicolumn{1}{c}{Value} \\
\hline                         
              &        \\
 Number of Channels (2 spares) & 40 \\
Input Signal & LVDS \\
Minimal Input Pulse Width & 5 ns \\
Channel Mask Register [+] & 40 bit \\
Time Window (min, max, step) [+]& 10, 150, 10 ns \\
Majority Trigger (low threshold) [+] & 1 to 39 channels\\
Majority Trigger (High Threshold) [+] & 1 to 39 channels \\
Trigger Bit Pattern & 40 bit \\
          &       \\
\hline
\end{tabular} 
\end{center}
\end{table}

 The master V1495 trigger module 
increments a trigger-number (Trigger-ID) for each processed
trigger. 
This Trigger-ID is sent to the veto system 
so that all veto events within the TPC acquisition window are marked with the 
same ID. The 
lower eight bits of the trigger number are sent to the V1720 as an 8-bit 
TPC trigger ordinal to further correlate an ADC event with the trigger.
There are also different Trigger-type identifiers as described in the 
next sections. 

\section{G2 Trigger}
      
\begin{figure}[h]
\begin{center}
\includegraphics[width = 9.cm ]{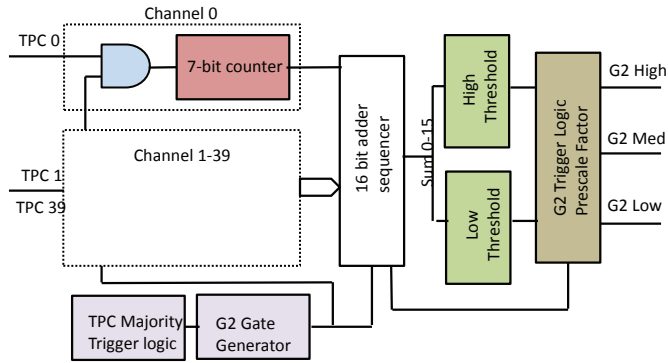} 
\caption{A block diagram of the G2 trigger logic}
\label{g2_trigger}
\end{center}
\end{figure}

 The G2 trigger provides a method to study signals and 
backgrounds within a selected energy range. As an example, it can
tag events with high-multiplicity S1 signals.  It is based on 
summing hits from  
all enabled TPC discriminator channels occurring within a fixed
time window which is opened after a pre-trigger is generated by a 
TPC majority trigger.  The trigger logic uses the 
same discriminator signals which are used to form the TPC majority 
trigger logic. A block-diagram of the G2 trigger logic is shown in
Figure ~\ref{g2_trigger}. The trigger logic follows the following steps. \\

\begin{itemize}
\item The trigger pattern is latched in an internal register after
 a positive TPC majority logic condition generates a pre-trigger.
\item A time window of 1 to 5~$\mu$s is immediately generated by 
a gate generator and used to inhibit other TPC pre-triggers.
\item  7-bit counters (one for each of the 38 channels 
plus 2 spares) start counting hits within the window.  The counters
  either stop at the end of the window, or stop when a counter reaches a
  maximum value of 127. 
\item After the window closes, a sequencer scans the counters, summing 
  their values. The TPC pre-triggers remain inhibited by an adder busy
  signal for an additional 0.85~$\mu$s.
\item The resulting sequencer sum is compared to two thresholds.
If the sum exceeds a high threshold a ``High'' multiplicity G2
pre-trigger is generated. If the sum is less than the high
threshold, but exceeds the low
threshold a ``Medium''  multiplicity G2 pre-trigger is generated. If
the sum is lesser than the Low threshold, a ``Low'' multiplicity
G2 pre-trigger is generated.
\item Both the High and Medium trigger types can be pre-scaled
  by separate, programmable factors.  The triggers are then provided to the 
system trigger logic and used in the trigger selection.
\item The process re-starts after the 
  pre-scale window closes and the counters are initialized. 
\end{itemize}

The dead time associated with any 
G2 trigger is equal to the sum of the time window, the 0.85~$\mu$
adder busy, and the G2 acquisition
window inhibit. The G2 acquisition-window-inhibit setting is equal to
the V1720 ADC acquisition window setting.\\  

\subsection{Veto Trigger}

The veto triggers are obtained in a PXIe-7961 
module ~\cite{NI7961} which is similar to the 
CAEN V1495 module, as both allow
trigger logic to be programmed into an on-board FPGA.
Veto trigger signals  are combined and sent to the  
 master V1495 located in the CLR.  
When a Veto trigger is received, 
all inputs to the V1495 are evaluated and if appropriate, a separate 
trigger signal, using Trigger-type bits, is stored in the Trigger-ID. Then a
trigger signal and the Trigger-ID are sent back to a  
CoR fanout and to   
each of the 4 PXIe timing modules in the 4 NI PXIe crates. The signals are 
finally distributed to all the PXIe modules through an internal bus. 
The PXIe timing modules in the crates also 
receive the $50$~MHz clock, and the 1PPS signal.
The veto trigger has the 2 operational modes given below. \\
\begin{enumerate}
\item {\bf TPC-only-mode}. In this configuration the trigger rate is determined 
by the internal activity of the TPC. The length of the data acquisition window 
is set to $45-70$~$\mu$s to detect prompt and delayed events
 in the veto  which are correlated to TPC activity
\item {\bf Pass-through-mode} In this configuration, the LSV 
and CTF triggers are logically combined (or) and used as the
input trigger to the master V1495. For this mode, the input trigger is 
determined by the LSV trigger rate. The length of the data 
acquisition window is set to $4.5$~$\mu$s in order to acquire the 
entire scintillation signal and must be adjusted to account for the delay 
in the  hardware trigger.
\end{enumerate}
 
The veto FPGA  has internal FIFO memory 
used to store the Trigger-ID for each incoming Veto trigger.  This
memory is read for each event to provide a trigger 
marker. The veto V1495 decodes the 
Trigger-ID and places it in the memory for readout via the VME 
bus for inclusion in the TPC data stream.   
A 1PPS signal from the LNGS GPS receiver is used to synchronize 
the V1495 and the veto FPGA modules.\\

\subsection{Synchronization}

\begin{figure}[h]
\begin{center}
\includegraphics[width = 15cm ]{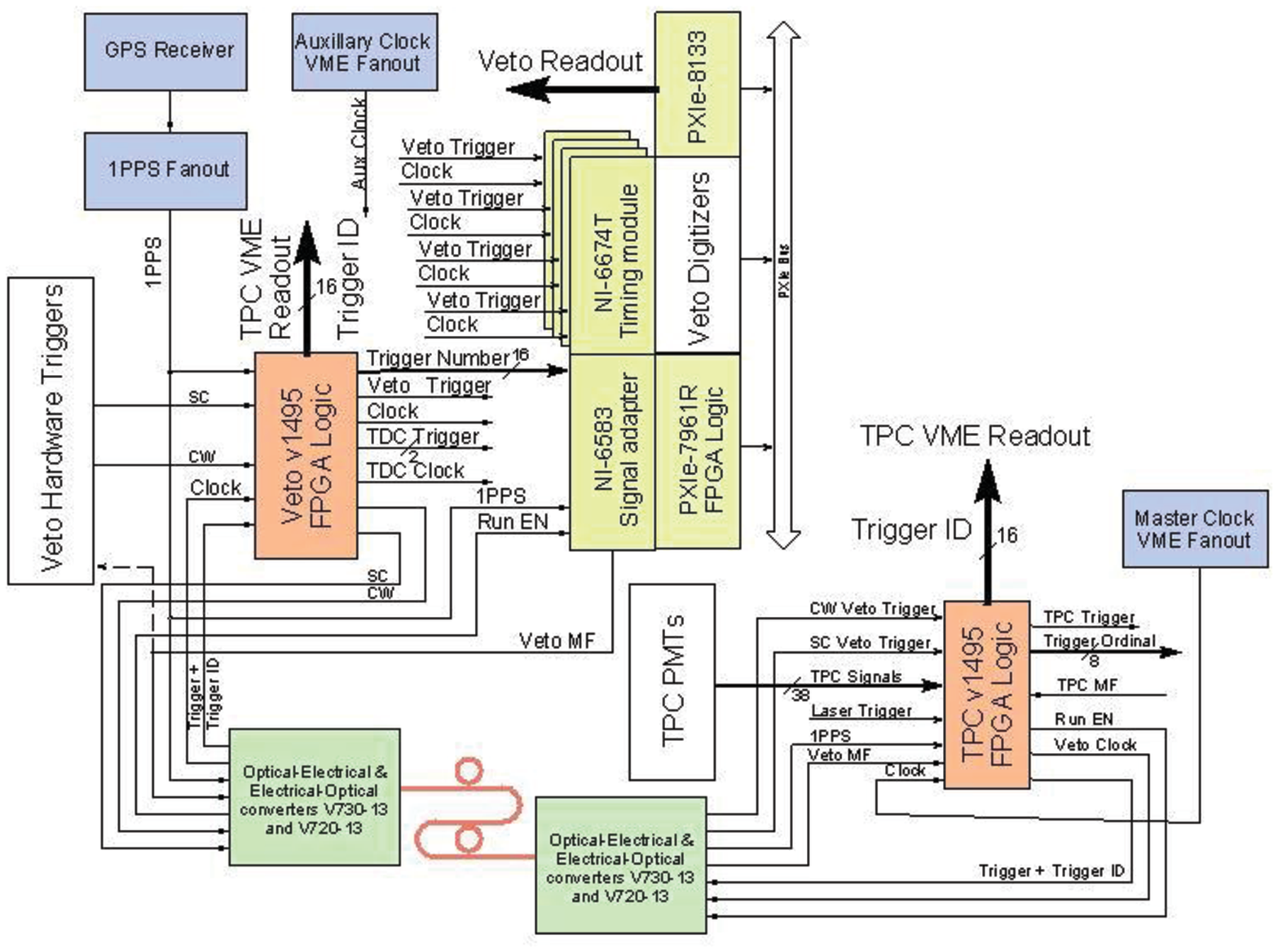} 
\caption{A block diagram showing the trigger connections
scintillation vetoes and the TPC cryostat}
\label{global_trigger}
\end{center}
\end{figure}

Figure ~\ref{global_trigger} shows the generation and flow of 
trigger signals. Unless otherwise noted, signals in the block diagram are 
active high TTL, ECL or LVDS. There are four FPGA-based trigger logic units. 
These are;
\begin{enumerate}
\item TPC V1495 trigger module located at the TPC site (CLR);
\item V1495 trigger module located at the Veto site (CoR);
\item Veto Fanout V1495 module located at the veto site (CoR);
\item Veto FPGA PXie-7961 logic module located in the PXIe crate (CoR)
\end{enumerate}

A 32-bit Trigger-ID is 
generated for each trigger, acquired by the PXIe timing module, 
and routed to the veto FPGA for decoding. The time stamps and the 
Trigger-ID are saved in a 
DMA buffer in the veto FPGA and are read asynchronously by the veto DAQ.
Thus, the Trigger-ID is read out via the VME bus as a part of 
the TPC data stream, and the
same Trigger-ID and the trigger signal are also available for the veto DAQ 
readout via a digital adapter module, NI 6583 ~\cite{NI6583},
attached to a veto FPGA. 
A 24-bit ``Event-counter'' is programmed to count all incoming triggers for each
event. This allows the same event to be correlated between different ADCs
as they all have identical time tags.  The event counters are stored in the 
data header. Both the TPC and veto counters are synchronized at the 
beginning of 
the run by the external Run-enable signal, and all systems are
kept in synchronization using a ``star'' distribution with equal time delays of
Clock, Trigger and Run-enable signals. The 
TPC V1495 trigger output is enabled at the same time as the Run-enable 
signal and the veto system receives Run-enable signal from the TPC
V1495 trigger. \\

It is necessary to inhibit 
triggers when any ADC memory is full. This is 
accomplished using a chained ``or'' of the ADC memory-full 
signals and firmware in 
the V1720 TPC ADCs. 
 Without these signals it would not be possible to guarantee event 
synchronization between two sub-systems. Generation of any 
Trigger-type is inhibited for an appropriate time after a trigger is
issued. \\

Three counters are recorded and used to create a time stamp for each trigger. 
These are listed below. 
\begin{itemize} 
\item The 1PPS counter counts the number of seconds elapsed 
between the trigger and Run-enable signal. The Run-enable signal 
occurs at run-start so this counter represents the number of seconds between 
the start of the run and the trigger.
\item The GPS fine-time counter counts the number of $50$~MHz clock cycles  
elapsed between the trigger and either the Run-enable or 1PPS signal,
as determined by whichever signal occurs last. The fine-time counter 
represents the number of 20 ~ns bins between either the Run-enable 
or 1PPS signal depending on the selected signal. 
\item The GPS one-second-counter counts the number of $50$~MHz clock 
cycles between the 1PPS signal occurring just prior to the 
Run-enable signal and the Run-enable signal itself. This counter 
is used to provide the number of $50$MHz cycles between a prior 1PPS 
signal and the following Run-enable signal. 
\end{itemize}

The counters are stored in the veto FPGA  memory, and
provide the time of a trigger with respect to the run-start
with a precision of $20$~ns.  This creates a unique time-stamp for 
a triggered event.  Identical time stamps are computed by
the TPC and veto DAQ systems, so that synchronization of events 
can be maintained.\\

\subsection{TDC}

A secondary data stream for the Veto system is implemented using TDCs. 
One of the discriminated FEDB veto outputs from each channel is connected
to a CAEN V1190 multi-hit 128 channel TDC ~\cite{caen1190}. The two 1190 
modules are located in the CoR and connect to the TPC V1495 trigger unit through
the local V1495 fanout logic unit. This secondary
system does not
replace the main veto readout but provides a back-up system and 
solves some TPC integration issues. Consequently
 the TDC provides an independent data stream for the LSV and CTF vetoes, and
is synchronized to the TPC trigger providing a check on the efficiency
of the main veto trigger. 
Thus a TPC event has direct mapping to the TDC data, and is correlated
to the veto ADC stream, solving a problem 
when the veto triggers before a TPC trigger is generated.  
\\

\subsection{Clock}

\begin{figure}[h]
\begin{center}
\includegraphics[width = 9.5cm ]{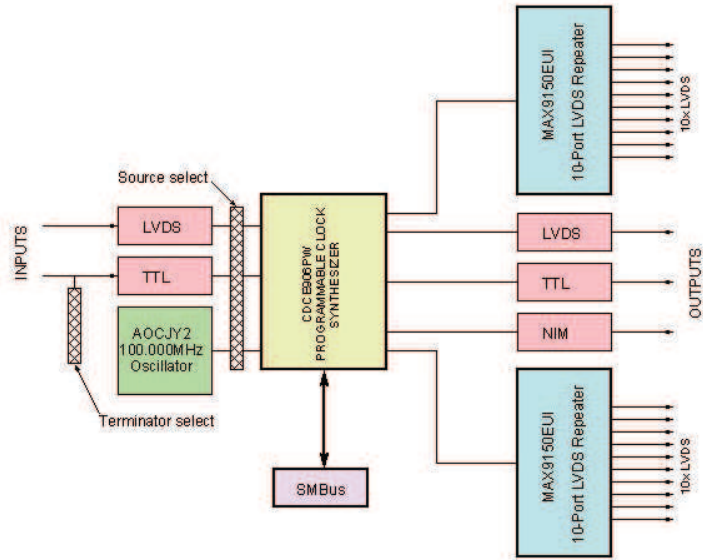} 
\caption{The block diagram of the CFM module}
\label{cfm_fig}
\end{center}
\end{figure}

The clock and fan-out, CFM ~\cite{baldin}, is an ``in-house'', single
width 6U VME module. 
The input accepts TTL, LVTTL, or CMOS signals with 
selective 50 or 10K $\Omega$ termination. In addition a second
connector accepts standard LVDS signals and is pin compatible with the
 CAEN V1720 ADC pinouts. The CFM uses a Texas Instruments CDCE906 
programmable 3-PLL clock synthesizer/multiplier/divider ~\cite{TIcdce906} 
and is the source of the output signals. The CDCE906 operates in various modes 
including ``PLL-bypass'' and ``Divider''. Jumper pins allows the input 
frequency from either from an external source  or
 from an on-board 100 MHz OCXO AOCJY2 ~\cite{ocxo} quartz crystal oscillator. 
 The block-diagram of the CFM design is shown in
Figure ~\ref{cfm_fig}. A 2x8-channel CAEN V976 fanout module is 
used to distribute Trigger and Run-enable signals. \\ 

\section{Data Acquisition}

It is assumed that the TPC and veto systems run 
independently at selected 
times, collecting data with their own triggers. For this reason, two modes of 
operation are available, which can be selected  using a run mode register 
stored at each site. \\
\begin{enumerate}
\item The system can run in local mode, in which each sub-system 
receives and records only its own triggers; or 
\item The system can run with event synchronization were the TPC DAQ system 
is the master. 
\end{enumerate}

\subsection{TPC DAQ}

\begin{figure}[h]
\begin{center}
\includegraphics[width = 12cm ]{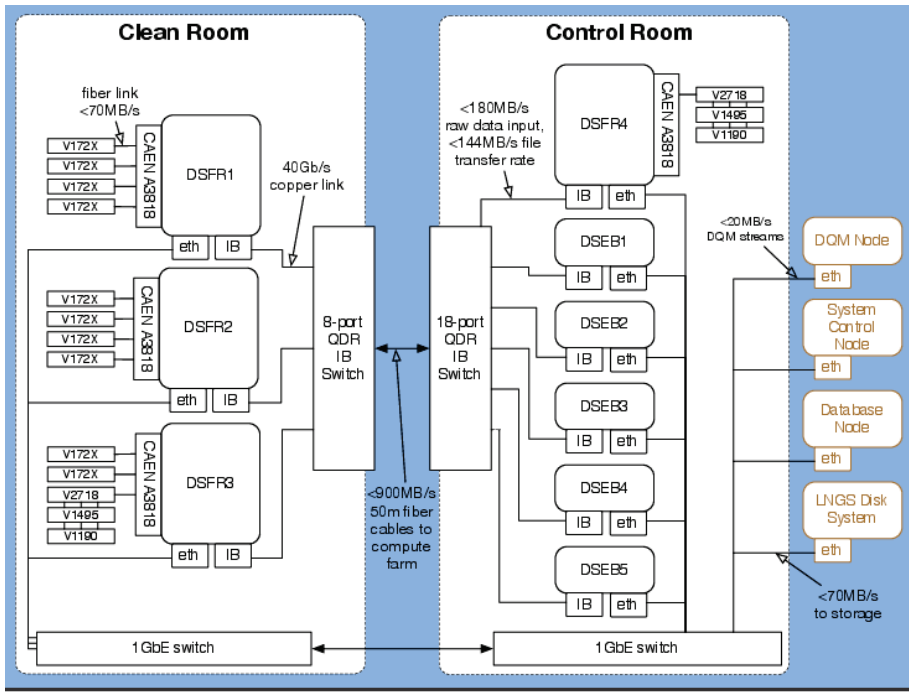} 
\caption{A Block diagram of the DAQ
illustrating the data flows between the CLR, CoR, Control, and Storage}
\label{tpc_diagram}
\end{center}
\end{figure}

\begin{figure}[h]
\begin{center}
\includegraphics[width = 14cm ]{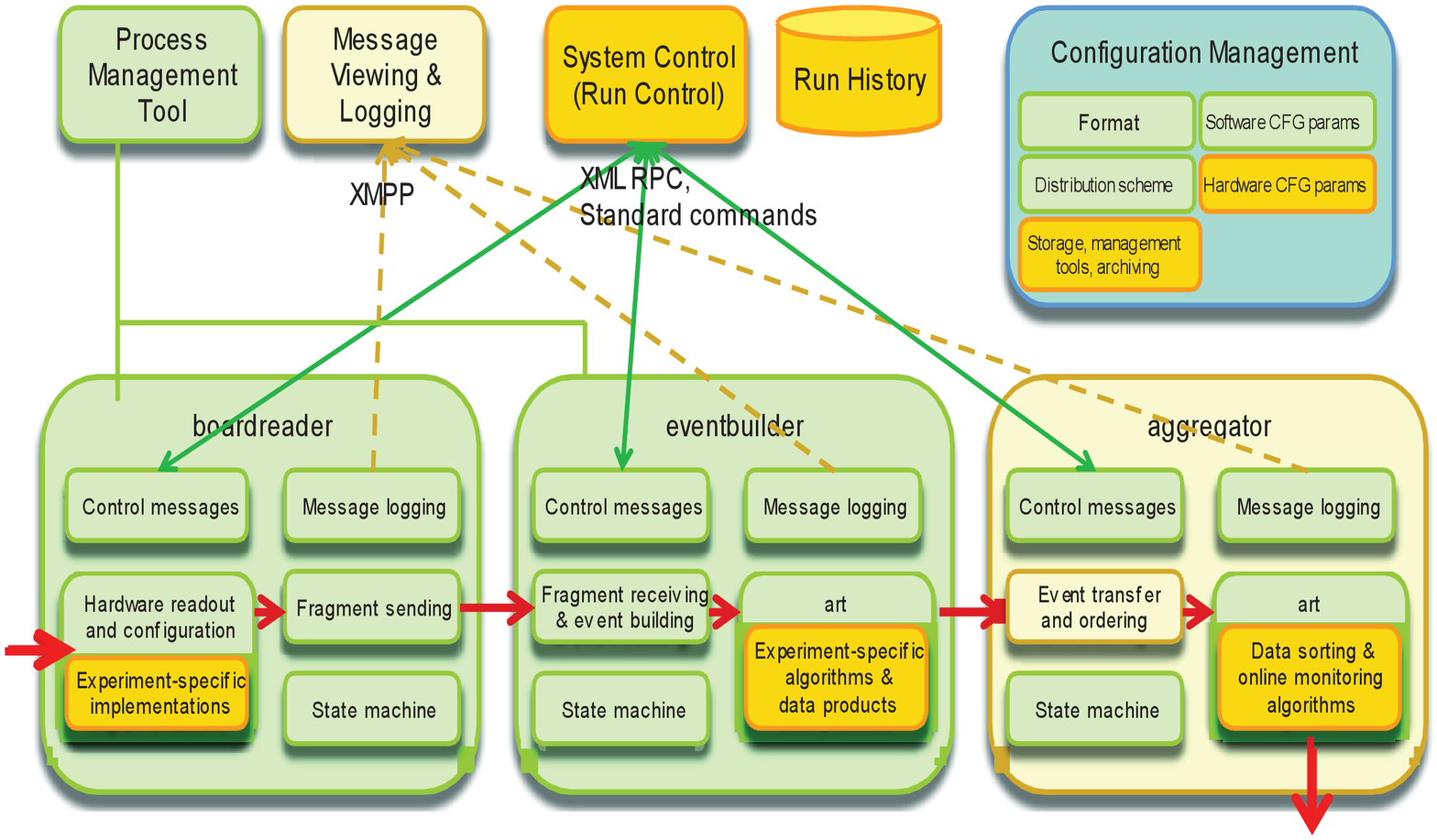} 
\caption{A Block Diagram of ArtDaq operations}
\label{artdaq}
\end{center}
\end{figure}

The DAQ is designed for high parallelization, as
illustrated in Figure ~\ref{tpc_diagram}. 
There are five, 8-channel CAEN V1724 digitizers and five, 8-channel
V1724 digitizers which are read via optical cable into the PCI
interface ~\cite{caenpci} of three local computers.  
One of these CLR computers is also optically connected to the 
VME bus through the CAEN 2718 crate controller, sharing information with 
the V1495 logic trigger module and therefore with the CAEN 1190 multi-hit TDCs 
in the CoR.
Digitizer output signals are sent to fragment servers and are multiplexed 
through a 40 Gb optical fiber from the 
CLR to the CoR. All computers run the ArtDaq software ~\cite{artdaq} tools 
which  transfer data
from the front-end computers through the DAQ system to disk, and 
provide a selection of events for online data quality monitoring, 
Figure ~\ref{artdaq}. 
The event builder compresses the data and an ``aggregator-server'' 
puts the events in sequential order storing them on disk. The
aggregator also provides events for online monitoring, including run 
parameters from disk files, and closes files at appropriate 
times during a run. \\

\subsection{Veto DAQ}

The veto DAQ architecture is designed to 
acquire data from 4 different PXIe
 controllers, preserving synchronization between each device and the TPC.
The system software is constructed of two software elements: 
acquisition-and-readout software and  builder software. 
Software is developed in LabVIEW, and conforms to LabVIEW coding 
and documentation standards.
The Veto DAQ uses the parallelism inherent in LabVIEW to perform the
main operations. \\

The readout software implements the following tasks:
\begin{itemize}
\item it acts as a server to listen for commands from the run
controller (\emph{e.g.} to request the  
start and stop data acquisition, communicate the status of the system, 
or provide  the number of acquired events);
 \item it checks  CPU, and memory
 usage in the controller;
\item it drives a state machine to perform the data acquisition operations 
({\it e.g.} initialize hardware, fetch data, stop, etc.);
\item it communicates with the FPGA module, retrieving the Trigger-ID and the 
time stamp of the event; 
\item it performs zero suppression on the acquired data;
\item it bundles zero-suppressed data from different digitizers in a
data fragment structure, which contains the time-stamps and the
Trigger-ID of the event; 
\item it sends data to the veto builder over the network;
\end{itemize}

Each veto readout software is installed in the PXIe controllers and is
automatically loaded when the controller boots. 
The veto DAQ architecture is based on a producer-consumer data flow
driven by a state machine. 
Communication over the network uses the LabVIEW Simple Messaging Reference
Library (STM). The content of a message is parsed, evaluated, and a
response message broadcast. 
During data acquisition, data fragments from every chassis are
collected by the veto builder. 
The builder checks the consistency of the Trigger-ID and timestamps of
the data fragments.  Finally, the events are entered in sequential
order and written on disk in a custom binary format. 

\section{Run Control}

 The DAQ sub-systems are handled by a common run controller which is
 configured to permit differential data acquisition modes. 
\begin{enumerate}
\item Global runs, where the TPC and veto triggers are correlated, and 
\item Local runs, where sub-detectors can be run independently.
\end{enumerate}

The run controller handles the initialization, start-up,
and stopping of data acquisition, and logs to the experimental data base 
relevant run  parameters, such as the TPC and Veto configuration. 
The run controller also supervises the data acquisition. When a 
sub-systems displays unusual behavior, data
acquisition is stopped and the DAQ sub-systems are reset and re-initialized.
The run controller is developed in LabVIEW, and conforms to LabVIEW coding 
and documentation standards.
Communication over a network with the TPC sub-systems is performed using the
XML-RPC protocol, while communication with Veto sub-systems is
performed using LabVIEW STM protocol.\\ 

\section{Summary}

This paper reports the electronics and data acquisition systems
developed for the DarkSide-50 DM search. The DAQ involves 
data collection from a 2-phase LAr TPC
which is enclosed by liquid scintillator and 
water Cherenkov vetos. The paper describes the electronics required to obtain
digitized PMT waveforms from the various detector systems, transmit
this data in parallel 
to event building computers, and maintain system 
synchronization. Event-building and run control are also discussed. 
 A dynamic range of over 2000 is obtained by using dual waveform 
digitizers for each TPC channel, and a novel cryogenic pre-amplifier 
was implemented which provides extremely low noise and signal
distortion. \\

Data acquisition
is stable with a maximum accepted data rate of over 50
Hz, and up to 6 TB of compressed data can be written to disk per
day.The full 
system has been operational several months, and can be  
controlled remotely. \\

\section{Acknowledgements}

This work was supported in part by grants from US National Science Foundation
 Nos. 1004051, 1242571, 1314268,, 1314501, 1314507, 1314479,
1314752, the US Department of Energy Nos. DE-FG02-91ER40671, DE-AC02CH11359
the Istituto Nazionale di Fisica Nucleare (Italy), and the National
Science Center Grant No. UMO-2012/05/E/ST2/20333 (Poland). The collaboration
gratefully acknowledges the support of  Nicola Cescato, Fabio Cortinovis, 
Thierry Debelle and Andrea Nobile of National Instruments. \\

\end{document}